# Building Data Science Capabilities into University Data Warehouse to Predict Graduation

Joonas Pesonen[1], Anna Fomkin[2], Lauri Jokipii[3]

[1]Rapida Ltd, Käpytie 18, 09220 Sammatti, Finland, joonas.pesonen@rapida.fi
[2]Aalto University, Otakaari 24, 02150 Espoo, Finland, anna.fomkin@aalto.fi
[3]Rapida Ltd, Käpytie 18, 09220 Sammatti, Finland, lauri.jokipii@rapida.fi

## 1. ABSTRACT

The discipline of data science emerged to combine statistical methods with computing. At Aalto University, Finland, we have taken first steps to bring educational data science as a part of daily operations of Management Information Services. This required changes in IT environment: we enhanced data warehouse infrastructure with a data science lab, where we can read predictive model training data from data warehouse database and use the created predictive models in database queries. We then conducted a data science pilot with an objective to predict students' graduation probability and time-to-degree with student registry data. Further ethical and legal considerations are needed before using predictions in daily operations of the university.

## 2. BACKGROUND

More and more data is generated daily in higher education. To use the growing amounts of data, we need IT infrastructure that supports different kinds of analytical activities. In the following we first define concepts related to data-driven decision making, then investigate the infrastructure and tools needed, and finally list requirements for a modern IT infrastructure for institutional analytics and educational data science.

### 2.1. Data Warehousing, Business Intelligence and Data Science

For a few decades, a *data warehouse* (DW) (e.g. Kimball & Ross 2011) has been the central information system supporting data-driven decision making in different organizations. A separate data warehouse populated with data pulled from different operational systems was needed since the data models allowing efficient transactional processing were not optimal for analytical needs. In higher education institutions, a typical data warehouse uses data from student information system (SIS), current research information system (CRIS), financial information system (FIS) and human resources information system (HRIS).

*Business intelligence & Analytics (BI&A)*, originating in business organizations, is referred to as the techniques, technologies, systems, practices, methodologies, and applications that analyze critical business data to help an enterprise better understand its business and market and make timely business decisions (Chen, Chiang & Storey 2012). There is an ever growing number of business intelligence applications helping users to analyse data and create interactive visualizations and dashboards. Methods in business intelligence are usually simple operations, such as cross tabulation, sums, averages and percentages.

For more advanced methods, such as statistical modeling and machine learning, specialized statistical software and/or use of a programming language such as R or Python is needed. Using these methods to support decision making has not traditionally been a part of business intelligence - the terms *advanced analytics* or *data science* have been used instead.

*Data science* was introduced by Cleveland (2001) as "a plan for expanding the technical areas of the field of statistics". He suggested that data science should be seen as an individual discipline instead of being a subdiscipline of computer science or statistics. The popularity of the term (and especially the term describing a practitioner of data science, a *data scientist*) has risen greatly in last few years, one example being Harvard Business Review article "Data Scientist: The Sexiest Job of the 21st Century" (Patil & Davenport 2012). Although it's hard to define comprehensively what data science is and what it is not, consensus is that data science combines domain knowledge, statistics and computer science (Figure 1).

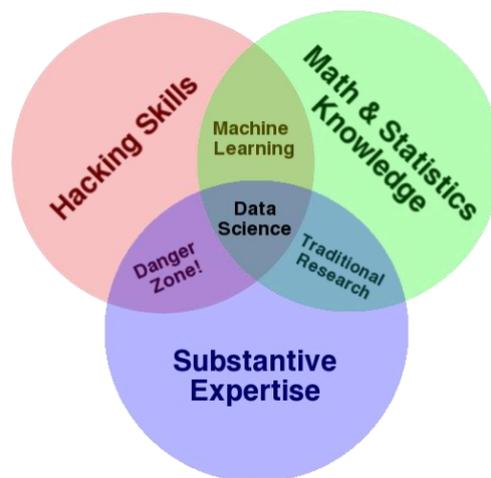

**Figure 1. The Data Science Venn Diagram (Conway 2010)**

## 2.2. Emerging Field of Educational Data Science

Systems and processes related to data-driven decision making are a growing area of interest in education. In literature, there are several distinct communities in this area, having different origins and focus. Piety, Hickey and Bishop (2014) recognise four distinct communities:

1. Academic/Institutional Analytics (higher education); originating in Institutional Research (IR; see e.g. Saupe 1990) community and having a focus on organizational development
2. Learning Analytics/Educational Data Mining; focusing on data about learning processes, and how to use that data to improve learning (see e.g. Siemens & Baker 2012)
3. Learner Analytics/Personalization; focusing on differences among learners and how they affect on student success
4. Systemic/Instructional Improvement (K12 and early childhood education); focusing on developing educational systems

Despite differences, there are many things common among these communities. Hence, Piety et al. (2014) suggest using *Educational Data Science* as a general term when dealing with questions of using analytical methods with educational data.

There are certain characteristics that distinguish educational data from other types of data (e.g. financial data) when considering analytics. Piety et al. (2014) identify four unique properties:

1. Human/social creation; data that is a product of human input may include errors
2. Measurement imprecision; measured things are not exact by nature
3. Comparability challenges; all aspects of education are not apparent in data
4. Fragmentation; ownership and governance of educational data is decentralized

One consequence of these properties is that an educational data scientist, besides statistics and programming skills, needs an understanding of educational sciences and the contexts and processes where the data is created (Buckingham, Hawksey, Baker, Jeffery, Behrens & Pea 2013). Availability of people with such a broad set of skills is a challenge, as it is difficult already to find people with general data science skills (Patil & Davenport 2012).

Although the concept of educational data science is useful, analytics in higher education is not only about educational data. A good example is *bibliometrics*, the discipline of analyzing written publications. Therefore, institutional research and institutional analytics remain as usable general concepts for describing different analytical activities carried out in higher education institutions.

## 2.3. IT Infrastructure for Institutional Analytics and Educational Data Science

There are many differences in data warehousing/business intelligence and data science approaches to data-driven decision making. In DW/BI, the aim is to produce a comprehensive system that allows several organization members to find insights from organization's data. Questions are often quite simple; for example "Which of our schools has best student retention rate?" or "Did we get better student feedback this year than last year?". Automatic procedures transfer the data regularly from the source systems, and making changes to the data warehouse needs coordination with several stakeholders.

In data science approach, a data scientist selects the data, tools and methods depending on the problem. Questions can be more complex; for example "Are there some specific courses causing trouble to students?" or "Which of our students are at risk to drop out?". The data scientist may use other data sources complementing data warehouse and build data models for the problem at hand, not trying to generate a general model for many use cases. This makes the development cycles fast compared to DW/BI approach.

We see that these approaches can complement each other. The data warehouse provides lots of interesting data for data science activities, and with data science it is possible to add new features to data warehouse, such as predictions, recommendations and alerts. However, for this to happen, the IT infrastructure for educational data science needs to be carefully planned and implemented. We see that in an optimal IT environment for educational data science there should be

- access to organizations own data (e.g. data warehouse)
- access to internet to get data from public sources
- several data processing, analysis and visualization tools to choose from
- a programming environment (e.g. R and Python)
- enough computing resources to do heavy calculations
- a process to integrate data science projects' results into existing data infrastructure (e.g. write predictive models into data warehouse)
- a process to ensure that data privacy is not violated

These aspects were considered when building data science capabilities into Aalto data warehouse infrastructure, further described in sections three and four.

## 3. INFORMATION PRODUCTION AT AALTO UNIVERSITY

In this section, we describe the organization information production at Aalto University and the IT infrastructure related to it.

## 3.1. Organisation of Information Production

Information production at Aalto University is operated in co-operation with different units. Management Information Services (MIS), a team in the Leadership Support Services of Aalto University, has a central role in information production. Team's objective is to support management in Aalto University and its units by providing up-to-date information about Aalto University's activities and outcomes. The most common end products are statistics, reports and analyses. A future goal is to focus more on predictive and prescriptive analytics.

Service areas of Management Information Services include:
- General information support for management
- Bibliometrics
- Student information statistics and reports
- Personnel reporting and services for Human Resources
- Government reporting and data for international rankings

Management Information Services participates in planning and development of university's information architecture and infrastructure and utilizes Aalto Data Warehouse. In information production, Management Information Services is a coordinating body receiving the end user needs and requirements and facilitating the collaboration needed between different parties (IT, learning services, research and innovation services, human resources and/or finance, depending on the case).

For example, student data is transferred to the Data Warehouse from Student Information System Oodi, operated by Learning Services of Aalto University. Changes in information production from Oodi to Data Warehouse and different data products are planned collaboratively. Different service areas may also have their own development projects directly with IT. If they later want to integrate these projects to the centralized information production, Management Information Services becomes involved.

### 3.2. Aalto University Data Warehouse

Aalto University's data warehouse consists of data from Student Information System, Current Research Information System, Financial Information Systems, Human Resources Information System and User Account Database. Data is used with several Business Intelligence tools, including ad-hoc analysis tools, visualization tools, formal report tools and dashboards. Data products created using these BI tools are disseminated in various forms, including interactive visualizations in a data portal and files sent via email. Some data is also published on university's public website.

As a part of Aalto University's goal to move towards predictive and prescriptive analytics, data science capabilities were added to data warehouse infrastructure. The new features were tested by conducting a data science pilot with an objective to predict students' graduation probability and time-to-degree with student registry data. This is further described in sections four and five.

## 4. DATA SCIENCE LAB AS A PART OF THE DATA WAREHOUSE

Data science capabilities were added to data warehouse infrastructure by creating a data science lab, an environment with data science tools (e.g. Anaconda, https://www.anaconda.com/) and a live connection to the data warehouse relational database. In the following, we first describe use of the data science lab and then consider the data privacy issues related to educational data science.

### 4.1. Educational Data Science Process

Data needed in statistical models and machine learning algorithms can be queried from data warehouse relational database with SQL. Additional data needed in data science activities can be imported as flat files or from APIs. An example would be labour market statistics fetched from statistics authorities. This model allows rapid iterations in data collection: as the data scientist gains more insight from the data, he/she can make instant alterations to data collection.

The products of educational data science activities include but are not limited to
- Prediction formulas obtained by fitting linear models (e.g. predicting graduation)
- Classifications of students, researchers, teachers and/or courses generated with clustering algorithms
- Keywords describing different textual artefacts (e.g. degree requirements, course descriptions) obtained with natural language processing
- Recommendations/matching (e.g. students and courses, students and thesis supervisors, students and research groups) generated with collaborative and/or content-based filtering

- Emotional tone classification of student feedback obtained with sentiment analysis
- Simulations, what-if-analysis

These products can be saved to the relational database, making them available in all tools that use data warehouse as a data source (Figure 2).

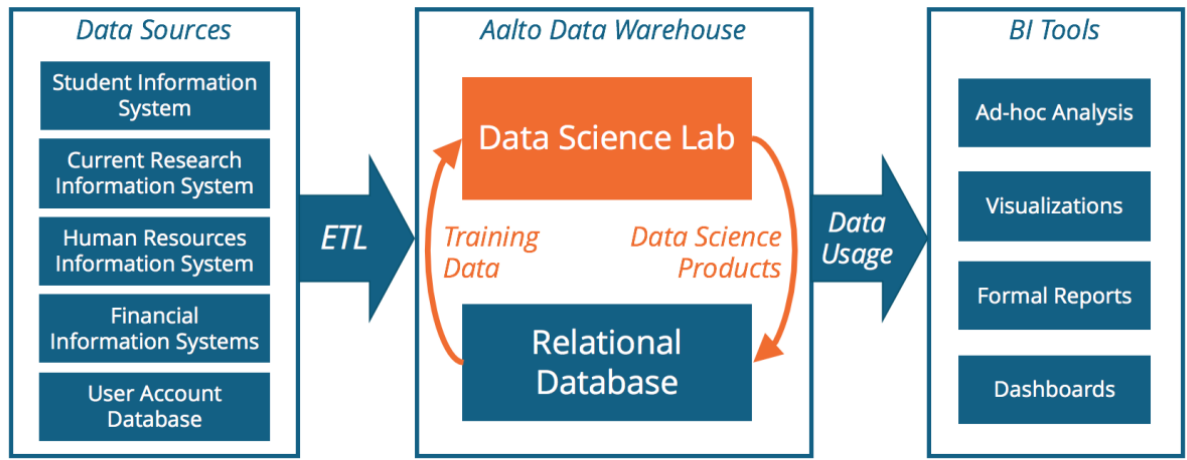

**Figure 2. Data Science -enabled Data Warehouse Infrastructure at Aalto University.**

## 4.2. Data Privacy and Educational Data Science

Privacy is often a concern with educational data. In our approach, the data stays in data warehouse infrastructure, and all the existing data and privacy policies can be used as such. However, extra care needs to be taken with educational data science activities, since it is likely that existing data and privacy policies do not cover novel use cases.

At the time of writing higher education institutions are preparing for the General Data Protection Regulation (2016/679) of the European Parliament and of the Council. One aspect that relates to GDPR and educational data science is student consent. Jisc (Sclater 2017) has suggested, that institutions should ask student consent for the use of sensitive data (e.g. ethnicity) and for taking interventions directly with the students based on analytics. They see that use of nonsensitive data for analytics can be considered as of *legitimate interest* or *public interest*, defined in GDPR.

We have followed this interpretation, thus excluding any sensitive data from our educational data science activities. Further ethical and legal considerations are also needed before taking any educational data science products into use on the level of individual students. However, many products can be used on group level (e.g. predicting the number of graduates of a specific study field during the next year) without privacy concerns.

## 5. PREDICTING GRADUATION PROBABILITY AND TIME-TO-DEGREE

The created data science environment was piloted with a goal to predict students' graduation probability and time-to-degree. The challenge in creating a model for time-to-degree is, how to handle students who are not likely to graduate at all. We followed a procedure, where logistic regression is first used to divide the student population in two groups based on whether they are likely to graduate within four years or not. Then linear regression is applied for the former group to predict their time-to-degree. For the other group, predicted time-to-degree is "four years or more".

In this section, we first describe the data collected to form the predictive models PM1, PM2 and PM3 (Table 1). Then we describe each model in detail. Last we describe how the models can be used in data warehouse.

Table 1. Summaries of the predictive models.

| Model | Description | Type | N | $R^2$ |
|---|---|---|---|---|
| PM1 | Probability of graduation | Logistic Regression | 8546 | 0.1456 |
| PM2 | Probability of graduation within four years | Logistic Regression | 10730 | 0.4156 |
| PM3 | Time-to-degree for a student likely to graduate in four years | Linear Regression | 4168 | 0.404 |

## 5.1. Data Collection

Data is collected by querying the data warehouse with SQL. The collected data consists of a set of variables linked to study rights, each row representing one study right (Table 2).

Table 2. Description of the variables used in different predictive models (PM1, PM2, PM3).

| Variable | Description |
|---|---|
| gender_female | Whether student is female (1) or not (0). |
| gender_male | Whether student is male (1) or not (0). |
| field_engineering | Whether study field is engineering (1) or not (0). |
| field_arts_and_design | Whether study field is arts and design (1) or not (0). |
| field_business | Whether study field is business (1) or not (0). |
| sum_of_cr | Sum of study credits (ECTS) on the observation date |
| no_credits_in_18m | Whether student has completed any studies during 18 months prior the observation date (0) or not (1). |
| distance_to_validity_end | Number of years between observation date and end of study right validity (7 years from beginning of studies) |
| graduated | Whether student has graduated (1) or not (0). |
| graduates_in_4y | Whether student graduated within four years from observation date (1) or not (0). |
| semesters_to_degree | Number of semesters between observation date and graduation date. Empty if student has not graduated. |

Study rights are limited to those that

- started after 1 August 2005 (this date divides study data significantly due to Bologna process)
- consist of rights for combined Bachelor's and Master's Degree (most common type of study right in Finland, targeted duration 3 years for Bachelor's Degree and then 2 years for Master's Degree; study right is valid for 7 years)
- were active during observation date.

The observation date is a query parameter which allows us to easily observe students' situation at different points of time. For example, if observation date is set to 1 August 2011, only study rights that were active on that date are collected and sum of study credits is counted from credits that were registered 1 August 2011 or before.

## 5.2. Predictive Model One (PM1): Probability of Graduation

For predicting students graduation probability, a predictive model was built using logistic regression. 1 August 2009 was used as the observation data, so that at the time of data collection all the study rights in the dataset have been active for at least eight years. Students, who had not graduated at this time, were interpreted as "not graduating". Although there might be cases where graduation happens after eight years, expected amount of these cases is so low that it should not cause significant bias in the model.

The model coefficients are described in Table 3. The greatest effect on probability of graduation was on student having no credits in the last 18 months, the probability being only 5 % for such students. Study field being arts and design or engineering and gender being male decreased the probability of graduation, whereas larger sum of credits and distance to the end of study right validity increased it.

## 5.3. Predictive Model Two (PM2): Probability of Graduation Within Four Years

The second model was built for classifying the students based on whether they are likely to graduate in following four years or not. 1 August 2013 was used as the observation date for collecting training data, so that it was known for all students if they graduated within four years from the observation date.

The model coefficients described in Table 3 are similar to PM1 as expected. Only notable differences are that field of study being engineering and distance to the end of study right validity were not significant predictors in PM2.

## 5.4. Predictive Model Three (PM3): Predicted Time-to-degree

A multiple linear regression was calculated to predict time-to-degree based on gender, study field, sum of credits and distance to study right end date. A significant regression equation was found with an $R^2$ of 0.404. Student's gender being male and study field being arts and design increased time-to-degree with about 0.3 semesters each. Sum of credits decreased time-to-degree so that each 100 completed credits caused about 1.3 semesters decrease. Distance to the end of study right validity increased time-to-degree, with each year of causing 0.4 semesters increase.

Table 3. Predictive model coefficients.

| Coefficients | PM1 | PM2 | PM3 |
|---|---|---|---|
| Constant | -0.8304 | -3.2970 | 6.6596 |
| gender_male | -0.2643 | -0.3869 | 0.3095 |
| field_arts_and_design | -1.0129 | -1.0453 | 0.3532 |
| field_engineering | -0.3026 | nonsignificant | nonsignificant |
| no_credits_in_18m | -2.7880 | -2.0927 | nonsignificant |
| sum_of_cr | 0.0101 | 0.0188 | -0.0132 |
| distance_to_validity_end | 0.1925 | nonsignificant | 0.3408 |

We compared predicted time-to-degree to actual time-to-degree in training data, to evaluate the accuracy of PM3 (Table 4). Only 19,5 % of predictions gave same time-to-degree when rounded to full semesters. On precision of +/- 1 semesters and +/- 2 semesters, the percentages were 54,9 % and 78,9 %, respectively.

Table 4. Prediction accuracy of PM3.

| Precision | Percentage with graduation predicted correctly |
|---|---|
| Same semester | 19,5 % |
| +/- 1 semester | 54,9 % |
| +/- 2 semesters | 78,9 % |

## 5.5. Using the Models in the Data Warehouse

To use the predictive models inside data, specific SQL-queries including calculated columns using model coefficients were written. These queries may be created as database views to enable using the predictions in different BI tools.

Following the first model (PM1), the probability $P_1$ of student $x$ to graduate is

$$P_1 = \frac{e^{-0.8304-0.2643x_1-1.0129x_2-0.3026x_3-2.788x_4+0.0101x_5+0.1925x_6}}{1 + e^{-0.8304-0.2643x_1-1.0129x_2-0.3026x_3-2.788x_4+0.0101x_5+0.1925x_6}}$$

where $x_1$ is gender_male, $x_2$ is field_arts_and_design, $x_3$ is field_engineering, $x_4$ is no_credits_in_18m, $x_5$ is sum_of_cr and $x_6$ is distance_to_validity_end for student $x$.

Following the second model (PM2), the probability $P_2$ of student $x$ to graduate within four years is

$$P_2 = \frac{e^{-3.297-0.3869x_1-1.0453x_2-2.0927x_4+0.0188x_5}}{1 + e^{-3.297-0.3869x_1-1.0453x_2-2.0927x_4+0.0188x_5}}$$

where $x_1$ is gender_male, $x_2$ is field_arts_and_design, $x_4$ is no_credits_in_18m and $x_5$ is sum_of_cr for student $x$.

Following the third model (PM3), time-to-degree $t_x$ in semesters for student $x$ who is likely to graduate in four years is

$$t_x = 6.6596 + 0.3095x_1 + 0.3532x_2 - 0.0132x_5 + 0.3408x_6$$

where $x_1$ is gender_male, $x_2$ is field_arts_and_design, $x_5$ is sum_of_cr and $x_6$ is distance_to_validity_end for student $x$.

## 6. DISCUSSION

In this section we elaborate our results, first on the created infrastructure and then on our pilot study on student graduation. After that we consider the ethical aspects of educational data science.

## 6.1. Data Science and Data Warehouse Infrastructure

Our first objective was to enhance Data Warehouse with Data Science capabilities. We developed a model and environment, which combines the DW/BI and data science approaches to data-driven decision making. The model and the environment were tested with a pilot project of predicting graduation.

The benefits portrayed in section 4.1 were realized during the pilot. Data collection with parameterised SQL queries allowed us to rapidly change the data used in predictive models. We could create dozens of iterations of training data within one day, while getting better insight about different factors' effect on student graduation.

For example, we first used "sum of credits last year" and "sum of credits two years ago" as predictors. When investigating the data in detail, we noticed that these variables were not normally distributed, because of a large amount of students with no credits in last two years. We created a dummy variable "no credits in last two years", which turned out to be a very significant predictor. Then we iterated with different timescales, ending up with "no credits in last 18 months".

In our experience, the infrastructure proposed in section 4 works well. However, data science activities require quite much effort, and expectations need to be managed. We suggest starting with small experiments and gradually scaling up.

## 6.2. Reflection on Graduation Prediction Pilot

Finnish educational system university education is funded by government and free for EU students. Long duration of studies has been seen as a problem, and policies have been made to restrict the maximum duration of studies (Aalto University 2018). If a student wants to continue studying after the maximum duration (seven years for Bachelor's degree and Master's degree combined), he/she needs to apply for an extension with a realistic plan for finishing studies.

Due to this process, the date seven years from the beginning of studies has a special meaning and must be taken into account with predictions. Another thing to consider is students' financial aid, for which students need to complete certain amount of studies (5 credits per month) to be eligible, and which ends in any case after a certain period (typically 50 months).

In our pilot about student graduation, we found that gender, field of study, sum of credits and not having any credits in last 18 months were all significant predictors for graduation probability and time-to-degree. Male students and arts and design students were less likely to graduate, and larger sum of credits increased probability. Most significant predictor was not having any credits during last 18 months. The results are in line with our experiences and previous research (e.g. Viitanen 2016).

The prediction accuracy presented in Table 3 sets some limitations to use of PM3. Essentially, the model is usable for use cases where precision of +/- 2 semesters is acceptable. Accuracy could probably be increased with adding more predictor variables and using more advanced data science methods, such as decision trees or neural networks (e.g. Herzog 2006).

An interesting finding is, that the significance of study field being engineering disappeared between PM1 and PM2. We suspect this to be due to different observation dates. Changes were made in engineering programmes between 2009 and 2013, and according to national statistics, the graduation rates on engineering fields of Aalto University have been constantly increasing since 2008, catching up with business students' graduation rates (Finnish Ministry of Education and Culture 2018).

## 6.3. Ethical Considerations on Educational Data Science

According to our results, predicting time-to-degree based on registry data is difficult. This is understandable - there are several real life human factors not visible in the registry data that may have an effect on graduation, such as financial status, family status and health. Probably much more accurate models could be produced by taking into account these kinds of factors, but there are many ethical and legal aspects to consider first.

On the ethical point of view, much depends on how the data and the models will be used and what purpose they serve. If we have a model, that can very accurately predict whether certain student will graduate or not, what do we choose to do? Do we offer extra help and support - or the opposite: draw off any support as the student is not going to graduate anyway?

Cathy O'Neil (2017) warns that algorithms and predictive models may unintentionally promote social injustice and discriminate minorities: we tend to think that algorithms are value-free and neutral in principle, but actually many moral choices are made when choosing which data is paid attention to and which is left out. She gives an example of a teacher who got fired because of an algorithm which rates teachers based on how much their students' test scores increased compared to previous year. Her class had high test scores previous year possibly due to cheating, and her inability to raise the scores got her a bad evaluation and led to her being fired. She could not challenge the evaluation, because neither she nor the school had access to the algorithm logic. (O'Neil 2017).

There are ways to mitigate the risks related to predictive models. O'Neil (2017) demands that the logic of algorithms must be published and recommends fairness audits for algorithms. In Europe, EU GDPR (2016/679) states that the data subject has a right not to be subject to a decision based on automated processing. This does not forbid all automatic processing, but makes consent of the subject necessary. For example, Jisc suggests that consent should be asked from students before using their data in interventions (Sclater 2017). According to GDPR, consent can also be withdrawn at any time. This will hopefully help create a fair culture of educational data science in Europe, with predictive models being used for helping instead of ranking people.

## 8. AUTHORS' BIOGRAPHIES

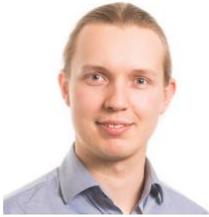
**Joonas Pesonen** received his Master's degree from University of Helsinki in 2013, majoring in mathematics with minors in education, psychology and computer science. During 2012-2017 he worked at CSC - IT Center for Science, building Finnish national education data infrastructure in close cooperation with Ministry of Education and National Agency for Education. In 2017 he co-founded Rapida Ltd, a company specializing in educational data science. His personal key competency areas are learning analytics, data visualization, data architecture and mathematical modeling.
joonas.pesonen@rapida.fi   https://www.linkedin.com/in/japesone/

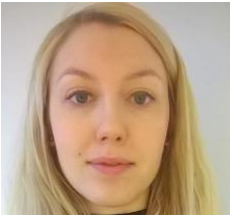
**Anna Fomkin** received her Master's degree from Aalto University in 2011, majoring in wood technology. She has been working in various positions in Learning Services at Aalto University and it's schools between 2009-2016 and has wide substantive expertise of student information data. Currently working as an information specialist in Management information Service at Aalto University.
anna.fomkin@aalto.fi   https://www.linkedin.com/in/anna-fomkin-38aa56114

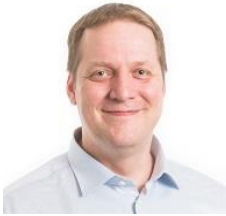
**Lauri Jokipii** studied Computer Science specializing in Information Systems and received his Master's degree from University of Helsinki in 2005. He is an experienced Senior Application Specialist with a demonstrated history of working in the information technology and services industry. Strong engineering professional skilled in Databases, Linux, Scripting, Data Warehousing, and Research. Working history includes working for University of Helsinki 2005-2011 and CSC - IT Center for Science 2011-2017. Currently Co-founder at Rapida Ltd.
lauri.jokipii@rapida.fi   https://www.linkedin.com/in/laurijokipii/